\pdfoutput=1
\documentclass[10pt]{article}
\usepackage{graphicx}
\def\Title#1{\begin{center} {\Large #1 } \end{center}}
\def\Author#1{\begin{center}{ \sc #1} \end{center}}
\def\Address#1{\begin{center}{ \it #1} \end{center}}

\newcommand\pubblock{\rightline{\begin{tabular}{l} Proceedings of the Second Annual LHCP \\ \pubnumber \\ \pubdate \end{tabular}}}
\newenvironment{Abstract}{\begin{quotation} \begin{center} \large ABSTRACT \end{center}\bigskip \begin{center} \begin{large}}{\end{large} \end{center} \end{quotation}}
\newenvironment{Presented}{\begin{quotation} \begin{center} PRESENTED AT \end{center}\bigskip \begin{center} \begin{large}}{\end{large} \end{center} \end{quotation}}

\def\beq{\begin{equation}}
\def\eeq#1{\label{#1}\end{equation}}
\def\eeqn{\end{equation}}
\def\beqa{\begin{eqnarray}}
\def\eeqa#1{\label{#1}\end{eqnarray}}
\def\eeqan{\end{eqnarray}}

\let\bar=\overbar

\def\Dslash{\not{\hbox{\kern-4pt $D$}}}
\def\dslash{\not{\hbox{\kern-2pt $\del$}}}

\def\msb{{\bar{\ssstyle M \kern -1pt S}}}

\newcommand{\lhc}{\mbox{LHC}\xspace}
\newcommand{\lhcb}{\mbox{LHCb}\xspace}
\newcommand{\qcd}{\mbox{QCD}\xspace}
\newcommand{\fsr}{\mbox{FSR}\xspace}
\newcommand{\nnlo}{\mbox{NNLO}\xspace}
\newcommand{\dynnlo}{\mbox{DYNNLO}\xspace}
\newcommand{\fewz}{\mbox{FEWZ}\xspace}
\newcommand{\pdf}{\mbox{PDF}\xspace}
\newcommand{\W}{\ensuremath{W}\xspace}
\newcommand{\Wp}{\ensuremath{W^{+}}\xspace}
\newcommand{\Wm}{\ensuremath{W^{-}}\xspace}
\newcommand{\Z}{\ensuremath{Z}\xspace}
\newcommand{\wmn}{\ensuremath{\W \to \mu\nu_{\mu}}\xspace}
\newcommand{\wpmn}{\ensuremath{\Wp \to \mu^{+}\nu_{\mu}}\xspace}
\newcommand{\wmmn}{\ensuremath{\Wm \to \mu^{-}\bar{\nu}_{\mu}}\xspace}
\newcommand{\wtn}{\ensuremath{\W \to \tau\nu_{\tau}}\xspace}
\newcommand{\zmm}{\ensuremath{\Z \to \mu\mu}\xspace}
\newcommand{\zee}{\ensuremath{\Z \to ee}\xspace}
\newcommand{\ztt}{\ensuremath{\Z \to \tau\tau}\xspace}
\newcommand{\D}{\ensuremath{D}\xspace}
\newcommand{\Dz}{\ensuremath{D^{0}}\xspace}
\newcommand{\Dp}{\ensuremath{D^{+}}\xspace}
\newcommand{\dzkp}{\ensuremath{\Dz \to K^{-}\pi^{+}}\xspace}
\newcommand{\dpkpp}{\ensuremath{\Dp \to K^{-}\pi^{+}\pi^{+}}\xspace}
\newcommand{\pp}{\ensuremath{pp}\xspace}
\newcommand{\pa}{\ensuremath{p{\rm Pb}}\xspace}
\newcommand{\sqs}{\ensuremath{\sqrt{s}}\xspace}
\newcommand{\pt}{\ensuremath{p_{\mathrm{T}}}\xspace}
\newcommand{\gevc}{\ensuremath{{\mathrm{\,Ge\kern -0.1em V\!/c}}}\xspace}
\newcommand{\gevcc}{\ensuremath{{\mathrm{\,Ge\kern -0.1em V\!/c}^{2}}}\xspace}
\newcommand{\tev}{\ensuremath{\mathrm{\,Te\kern -0.1em V}}\xspace}
\newcommand{\nb}{\ensuremath{{\mathrm{\,nb\!}}}\xspace}
\newcommand{\pb}{\ensuremath{{\mathrm{\,pb\!}}}\xspace}
\newcommand{\invnb}{\ensuremath{{\mathrm{\,nb\!}^{-1}}}\xspace}
\newcommand{\invpb}{\ensuremath{{\mathrm{\,pb\!}^{-1}}}\xspace}
\newcommand{\invfb}{\ensuremath{{\mathrm{\,fb\!}^{-1}}}\xspace}

\usepackage{color}
\usepackage{xspace}
\usepackage{lineno}
\usepackage{multirow}
\newcommand\pubnumber{LHCb-PROC-2014-020}
\newcommand\pubdate{\today}
\def\affiliation{On behalf of the \lhcb collaboration, \\
University of Birmingham, School of Physics and Astronomy \\
Edgbaston, Birmingham B15 2TT (United Kingdom)}
\begin{document}
\large
\begin{titlepage}
\pubblock

\vfill
\Title{Electroweak physics at \lhcb}
\vfill

\Author{Simone Bifani}
\Address{\affiliation}
\vfill

\begin{Abstract}
Measurements of electroweak boson production provide an important test of the Standard Model at the \lhc energies and allow the partonic content of the proton to be constrained. \W and \Z bosons are reconstructed in several leptonic final states using data samples corresponding to an integrated luminosity of up to about 1\invfb. Inclusive and associated production cross-sections are reported.
\end{Abstract}
\vfill

\begin{Presented}
The Second Annual Conference \\
 on Large Hadron Collider Physics \\
Columbia University, New York, U.S.A \\ 
June 2-7, 2014
\end{Presented}
\vfill
\end{titlepage}
\def\thefootnote{\fnsymbol{footnote}}
\setcounter{footnote}{0}
\normalsize 

\section{Introduction}

The \lhcb detector~\cite{lhcb} is a single-arm forward spectrometer covering the pseudorapidity range $2<\eta <5$, designed predominantly for the study of particles containing $b$ or $c$ quarks. The forward reach of the experiment can provide precise tests of the Standard Model at the \lhc energies in a kinematic range that is complementary to the General Purpose Detectors. Theoretical predictions of electroweak boson production, which are available at next-to-next-to-leading order (\nnlo) in perturbative quantum chromodynamics, rely on the parameterisations of the momentum fraction, Bjorken-$x$, of the partons inside the colliding particles. \lhcb can directly access the low Bjorken-$x$ region of the phase space, extending as low as $10^{-6}$, where the uncertainties on the parton density functions (\pdf) are larger, as well as probe high values.

Measurements based on data collected in \pp collisions at a centre-of-mass energy of \sqs = 7\tev corresponding to integrated luminosities between 37\invpb and 1\invfb are reported. Cross-sections for inclusive \W and \Z boson production are determined by reconstructing decays to final states with leptons. Studies of associated production of \Z bosons with jets and \D mesons, and the first observation of \Z production in proton-lead collisions at a centre-of-mass energy per proton-nucleon pair of $\sqrt{s_{NN}} = 5\tev$ using data corresponding to 1.6\invnb are also presented.

\section{Inclusive production}

\subsection{\wmn}

\W candidates are reconstructed via the \wmn final state by requiring an isolated muon with a transverse momentum, \pt, greater than 20\gevc and lying in the pseudorapidity range between 2.0 and 4.5~\cite{wz}. Additional criteria on consistency with the primary vertex and event activity are imposed to further reduce the background contamination. The signal purity is determined by simultaneously fitting the \pt spectra of positively and negatively charged muons in data to the expected shapes for signal (simulation) and background (simulation and data) contributions in five bins of muon pseudorapidity. Backgrounds that are accounted for in the fit include \zmm decays where one of the two muons goes outside the detector acceptance, \wtn and \ztt processes where one tau decays leptonically to a muon inside \lhcb, semileptonic decays of heavy flavour hadrons, and \qcd events where kaons or pions decay in flight to muons. The result of the fit is shown in Figure~\ref{fig:WZ} (left). A total of 14\hspace{0.5mm}660 \Wp and 11\hspace{0.5mm}618 \Wm events are selected, with a purity of $78.8\%$ and $78.4\%$, respectively.

\subsection{\zmm}

\zmm candidates are identified by requiring two reconstructed muons with opposite charge, $\pt > 20\gevc$, $2.0 < \eta < 4.5$ and having a combined invariant mass in the range $60 < M_{\mu\mu} < 120\gevcc$~\cite{zmm}. Data-driven methods are used to estimate backgrounds due to semileptonic decays of heavy flavour hadrons and \qcd events where kaons or pions either decay in flight or punch through the detector to be falsely identified as muons. Simulation is used to study contribution from \ztt, top and di-boson production. The selection retains \mbox{52\hspace{0.5mm}626} candidates, with a purity of $99.7 \%$. The distribution of the reconstructed di-muon invariant mass is presented in Figure~\ref{fig:WZ} (right).

The same approach is adopted to reconstruct \Z bosons in proton-lead collisions~\cite{zmmpa}. The data are used to study contamination due to muon misidentification and heavy flavour mesons decaying semileptonically. In total, 15 candidates are identified with a purity consistent with the level observed in \pp collisions.

\subsection{\zee}

Two identified and oppositely charged electrons with a \pt greater than $20\gevc$ and a pseudorapidity between 2.0 and 4.5 are combined to reconstruct \zee candidates~\cite{zee}. The electron momentum, rather than its energy deposition in the electromagnetic calorimeter, is used to reconstruct the two-body invariant mass, $M_{ee}$, which is required to be greater than 40\gevcc. Due to saturation of the calorimeter and an incomplete Bremsstrahlung recovery, the $M_{ee}$ distribution is significantly broader than the $M_{\mu\mu}$ peak. Background due to particle misidentification is estimated using a data sample of same-sign electron pairs. Residual contribution from heavy flavour hadrons and \ztt is found to be negligible. A total of 21\hspace{0.5mm}420 candidates are selected with a purity of $95.5\%$. The reconstructed di-electron invariant mass distribution is shown in Figure~\ref{fig:Z} (left).

\subsection{\ztt}

\ztt candidates are reconstructed via five final states, which are identified by different combinations of leptonic and semileptonic secondary tau decays ($\mu\mu$, $\mu e$, $e \mu$, $\mu h$ and $e h$)~\cite{ztt}. The leading lepton must have a transverse momentum exceeding 20\gevc, while the second particle is required to have $\pt > 10\gevc$. Leptons and hadrons are required to have a pseudorapidity in the range $2.0 < \eta < 4.5$ and $2.25 < \eta < 3.75$, respectively. The invariant mass of the visible particles must be above 20\gevcc. Further requirements are applied in order to reduce backgrounds: particles must be isolated and back-to-back in the transverse plane; tracks are required to be inconsistent with production at the primary vertex for the $\mu\mu$, $\mu h$ and $e h$ cases; due to a larger background from Drell-Yan events in the $\mu\mu$ final state, tracks are required not to be balanced in \pt. In total, 990 events are selected with a purity that ranges between 60\% and 70\% depending on the final state. The invariant mass distribution for the $\mu\mu$ final state is presented in Figure~\ref{fig:Z} (right).

\section{Associated production}

\subsection{$\zmm+jet$}

The \zmm analysis described above is adopted as the baseline selection to study jet production in association with a \Z boson~\cite{zmmj}. The anti-$k_{T}$ algorithm with a radius parameter of 0.5 is used to cluster jets in the pseudorapidity range between 2.0 and 4.5. Reconstructed tracks and energy deposits in the electromagnetic and hadronic calorimeters serve as charged and neutral particle inputs to the jet reconstruction. Each jet is required to be well separated from the decay products of the \Z. The jet energy is calibrated using simulation and checked in data, and the corresponding resolution is $10\mbox{-}15\%$ for jet transverse momenta between 10 and 100\gevc. The fraction of \zmm events with at least one jet with $\pt > 10(20)\gevc$ is determined to be 19.9(7.7)\% with a background contamination that is consistent with the inclusive \Z analysis.

\subsection{$\zmm+\D$}

\begin{figure}[!h]
\begin{center}
\includegraphics[height=.3\textwidth]{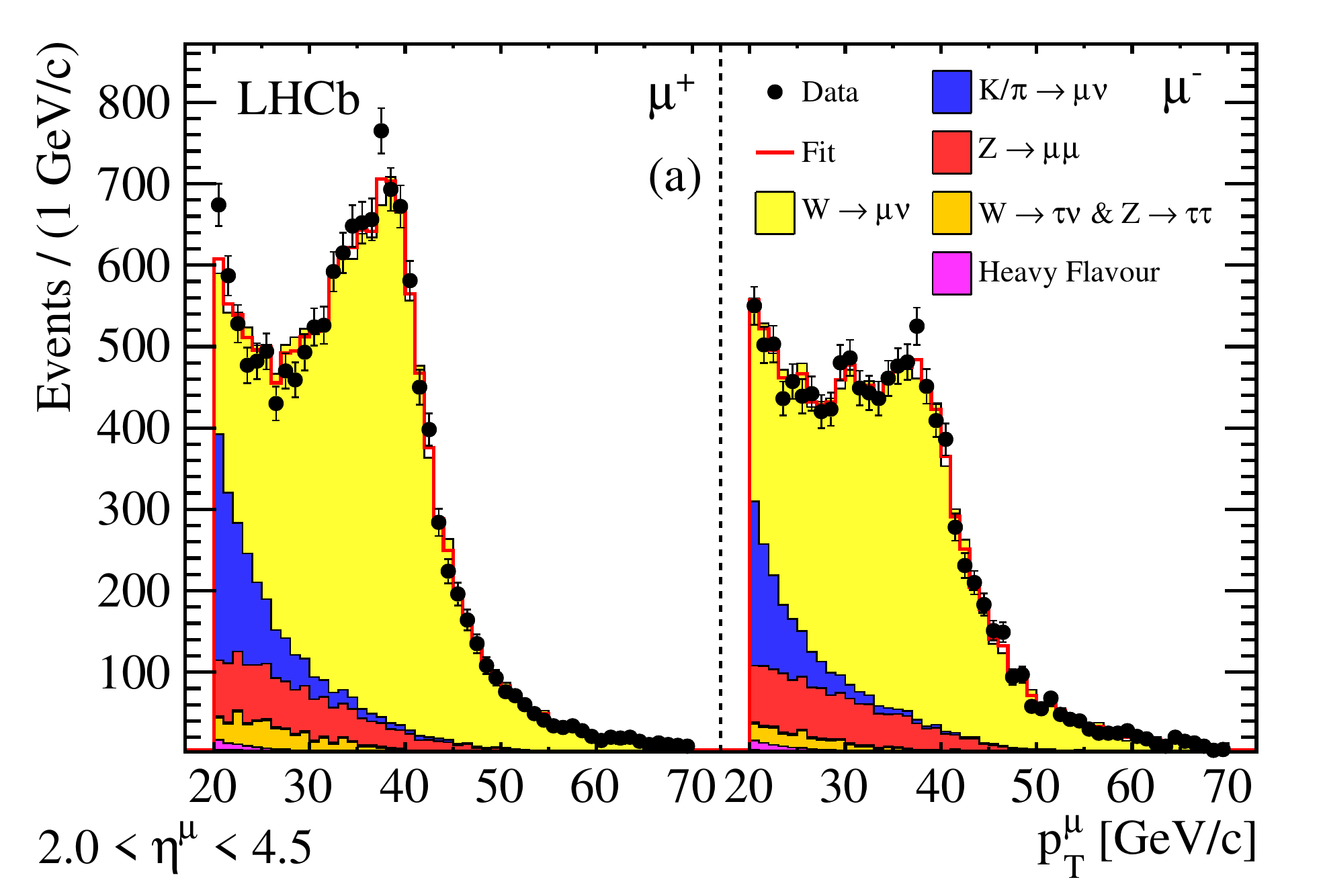}
\hspace{.5cm}
\includegraphics[height=.3\textwidth]{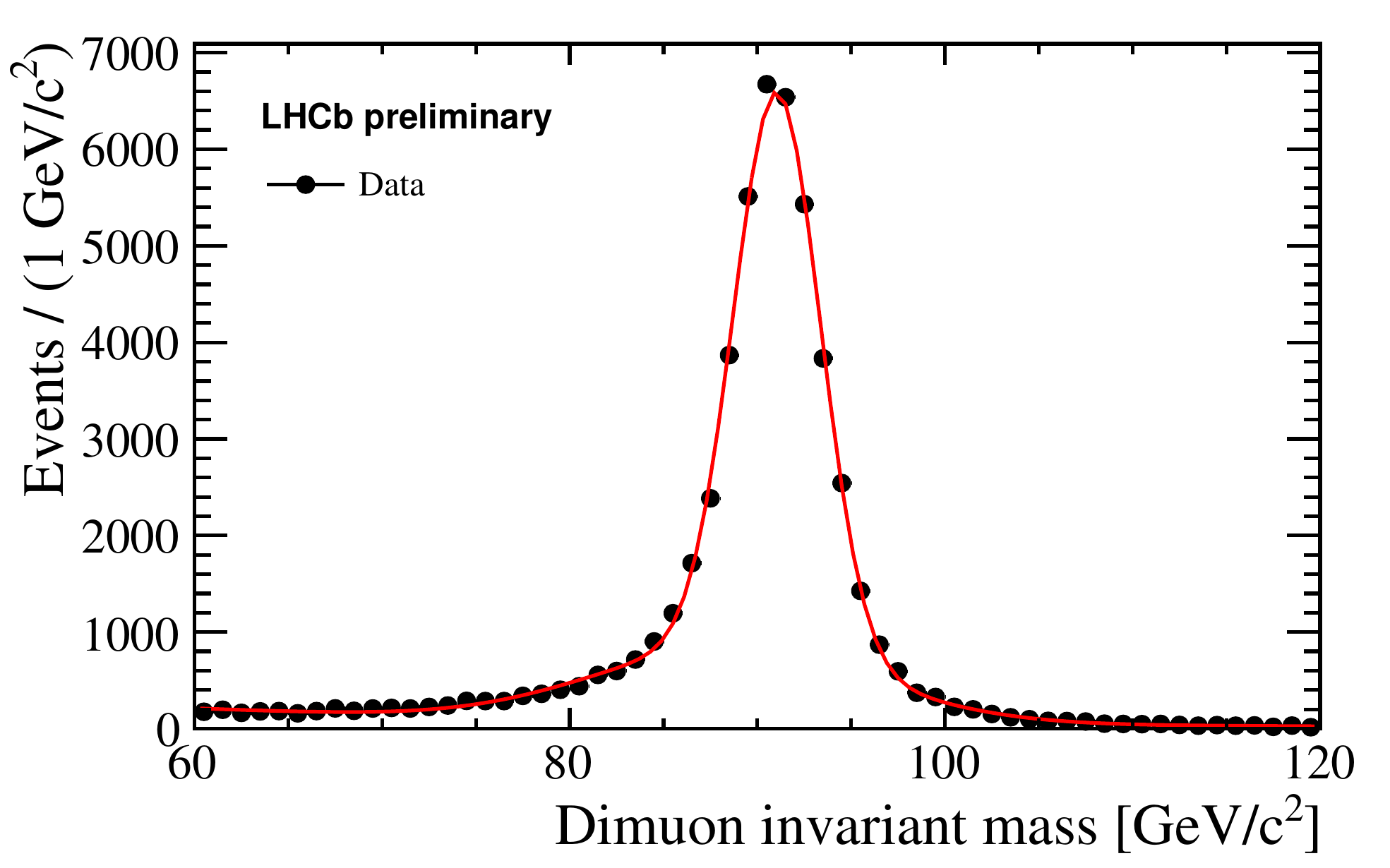}
\caption{Muon transverse momentum distribution of \wmn candidate events~\cite{wz} (left) and di-lepton invariant mass distribution for the \zmm final state~\cite{zmm} (right).}
\label{fig:WZ}
\end{center}

\vspace{.6cm}

\begin{center}
\includegraphics[height=.3\textwidth]{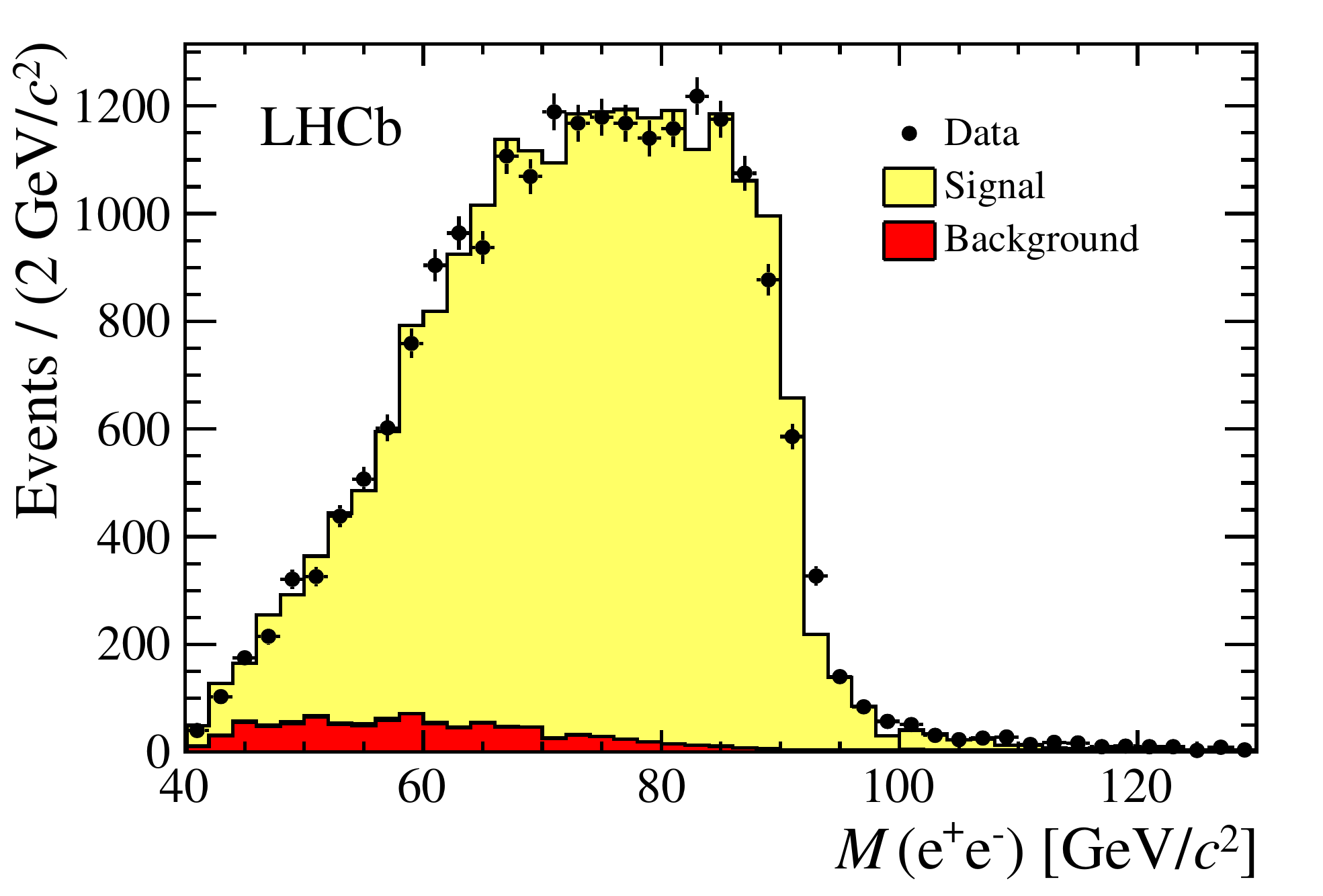}
\hspace{2.cm}
\includegraphics[height=.3\textwidth]{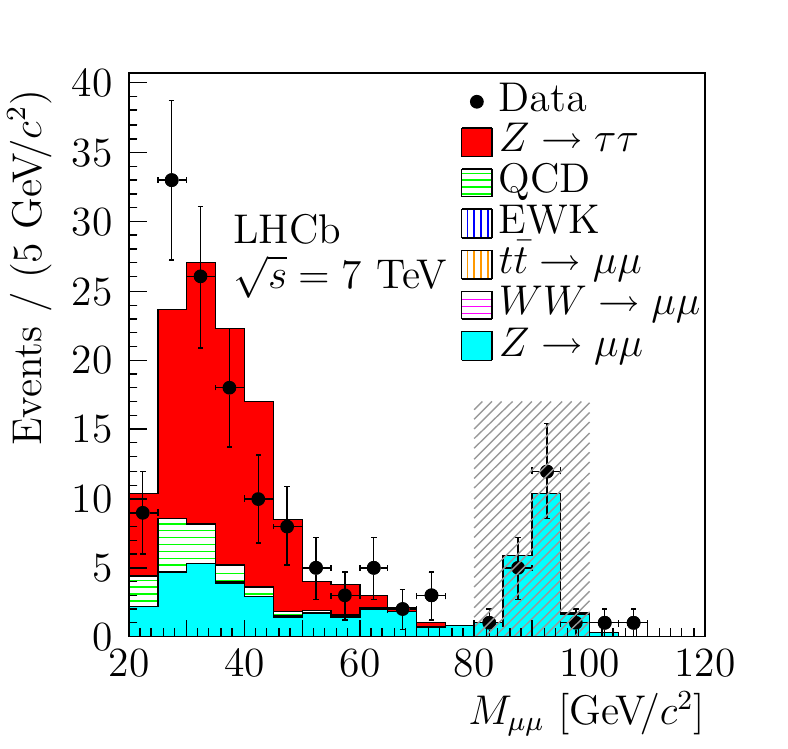}
\hspace{1.cm}
\caption{Di-lepton invariant mass distribution for the \zee~\cite{zee} (left) and \ztt~\cite{ztt} (right) final states.}
\label{fig:Z}
\end{center}

\vspace{.6cm}

\begin{center}
\includegraphics[height=.3\textwidth]{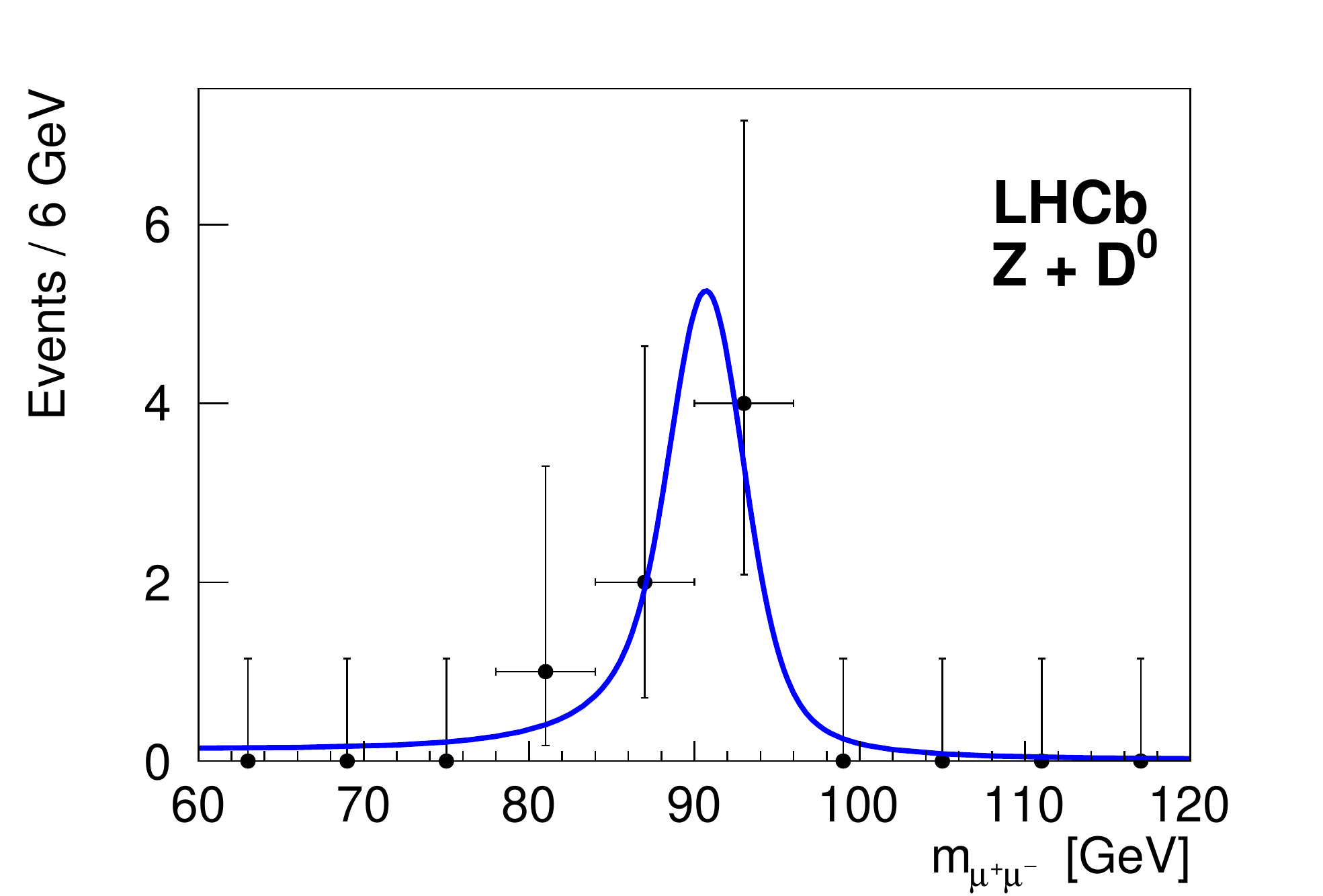}
\hspace{.5cm}
\includegraphics[height=.3\textwidth]{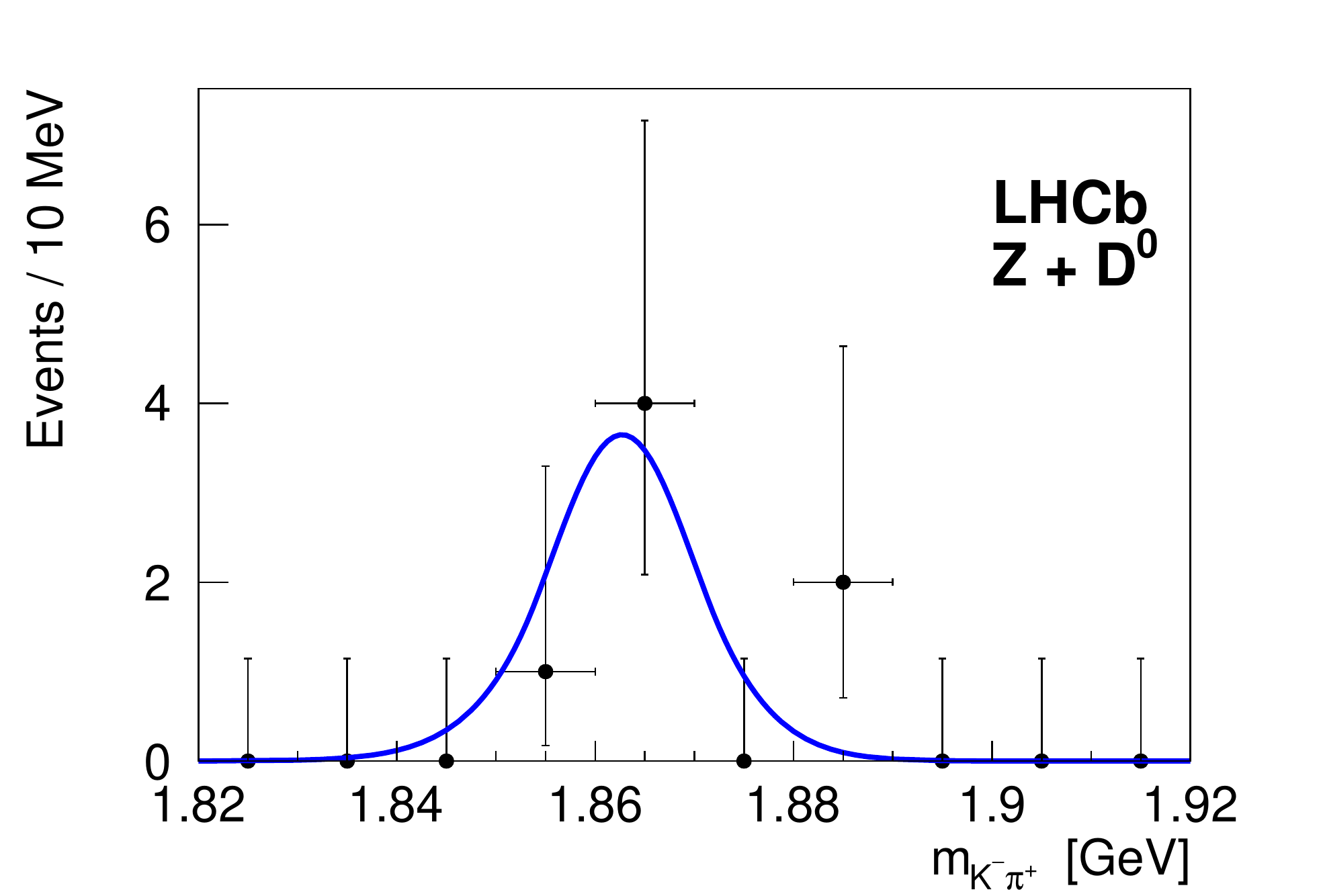}
\caption{Invariant mass distribution for \Z (left) and \D (right) candidates for $\Z+\Dz$ events~\cite{zmmd}.}
\label{fig:ZD}
\end{center}
\end{figure}

The standard \zmm selection is furthermore applied to search for associated production of a \Z boson with an open charm meson~\cite{zmmd}. \D mesons are reconstructed via \dzkp and \dpkpp decays in the invariant mass ranges $1.82 < M_{K^{-}\pi^{+}} < 1.92\gevcc$ and $1.82 < M_{K^{-}\pi^{+}\pi^{+}} < 1.91\gevcc$. Only charm mesons with $2 < y_{\D} < 4$ and $2 < p_{\mathrm{T},\D} < 12\gevc$ that are consistent with being produced at the same primary vertex as the \Z boson are considered. Seven candidates for $\Z+\Dz$ and four candidates for $\Z+\Dp$ associated production are observed. The combined significance of the observation is 5.1 standard deviations. The reconstructed \Z and \Dz invariant masses are shown in Figure~\ref{fig:ZD}.

\section{Results and conclusions}

Production cross-sections of electroweak bosons are measured with the kinematic requirements that leptons from the boson decay have a transverse momentum greater than 20\gevc, a pseudorapidity between 2.0 and 4.5, and, in case of \Z production, a combined invariant mass between 60 and $120\gevcc$. For associated \Z production with jets, two \pt thresholds of the jet are considered, 10 and 20\gevc. In case of $\Z+\D$ events, the charm meson is required to have $2 < y_{\D} < 4$ and $2 < p_{\mathrm{T},\D} < 12\gevc$. Results are corrected for trigger, track finding, particle identification and selection efficiencies, acceptance and loss due to final state radiation (\fsr). Efficiencies are primarily estimated using data-driven methods, while the acceptance and \fsr corrections are determined from simulation. The cross-section measurements are listed in Table~\ref{tab:cs}.

\begin{figure}[!h]
\begin{center}
\includegraphics[height=.31\textwidth]{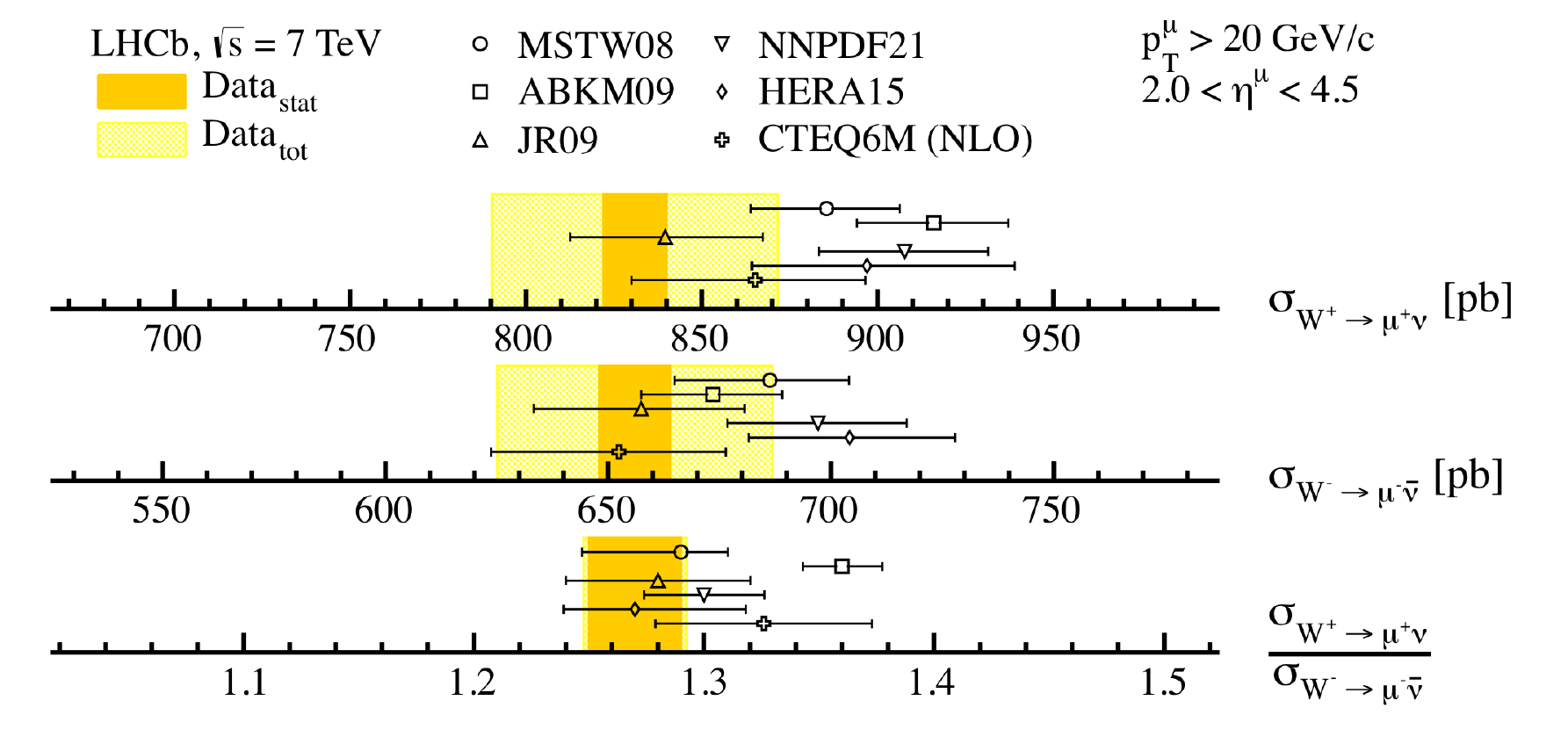}
\caption{Inclusive \W cross-section measurements compared to theoretical predictions~\cite{wz}.}
\label{fig:CSW}
\end{center}

\vspace{.6cm}

\begin{center}
\includegraphics[height=.26\textwidth]{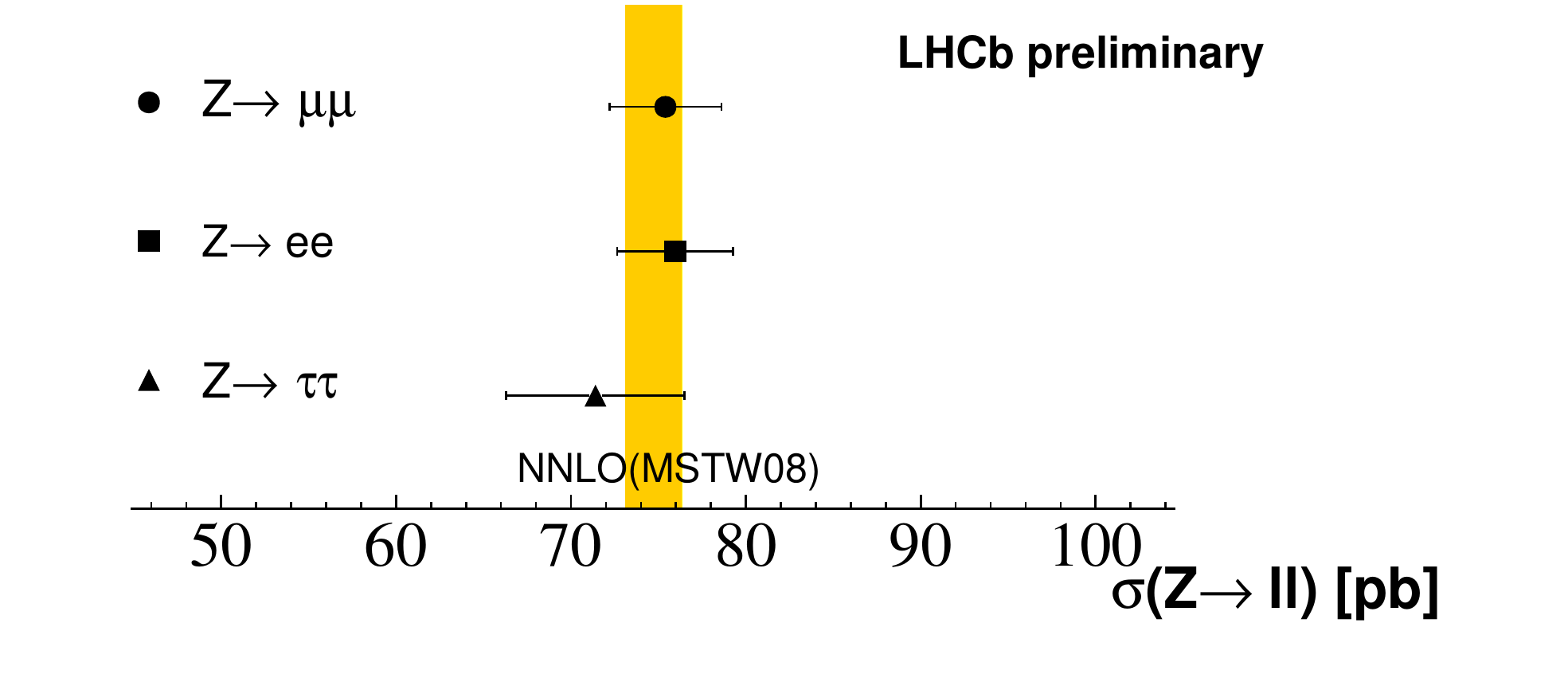}
\includegraphics[height=.26\textwidth]{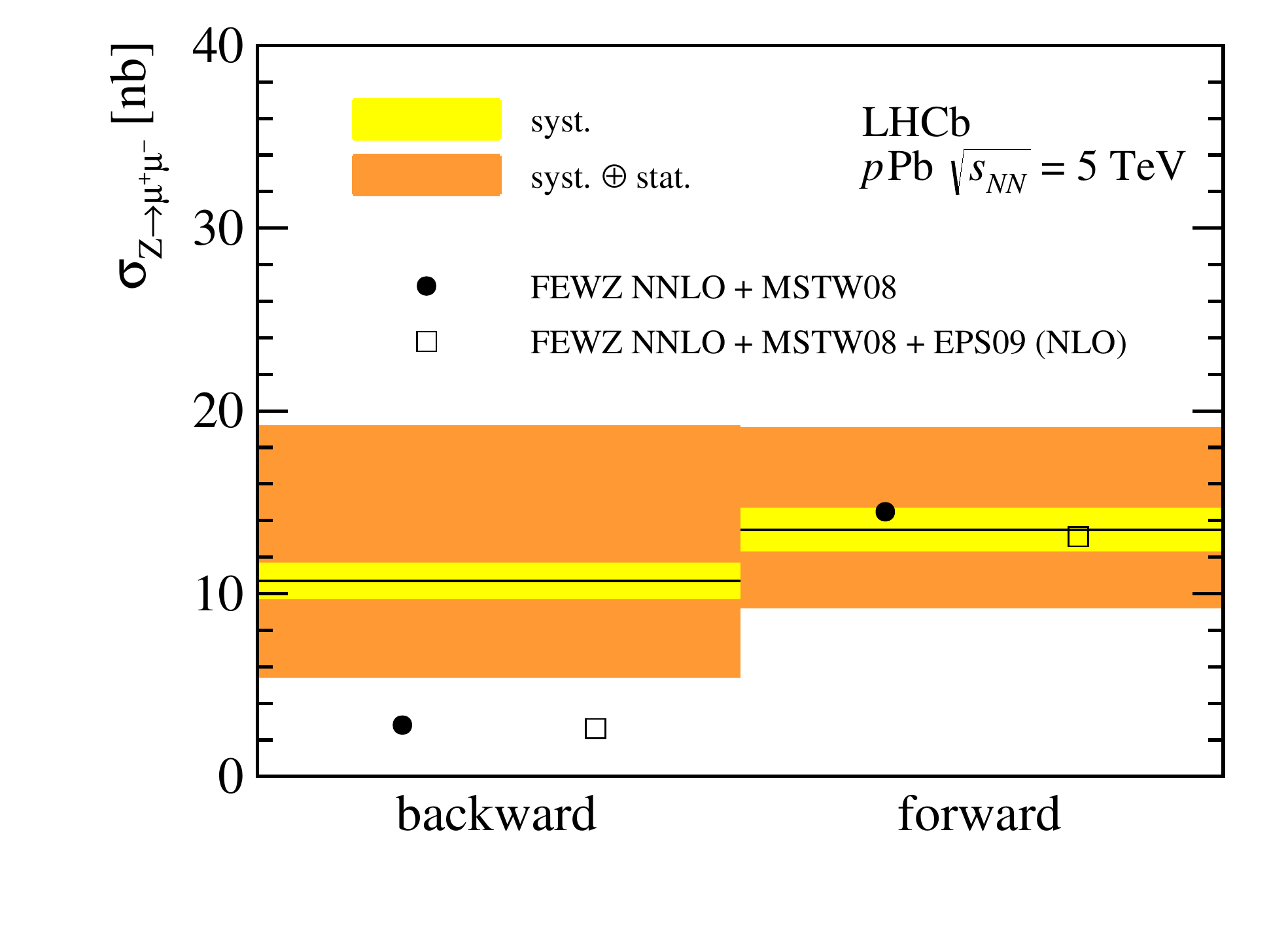}
\caption{Inclusive \Z cross-section measurements in \pp~\cite{zmm} (left) and \pa~\cite{zmmpa} (right) collisions compared to theoretical predictions. `Forward' (`backward') refers to positive (negative) rapidity values defined relative to the direction of the proton beam.}
\label{fig:CSZll}
\end{center}

\vspace{.6cm}

\begin{center}
\includegraphics[height=.36\textwidth]{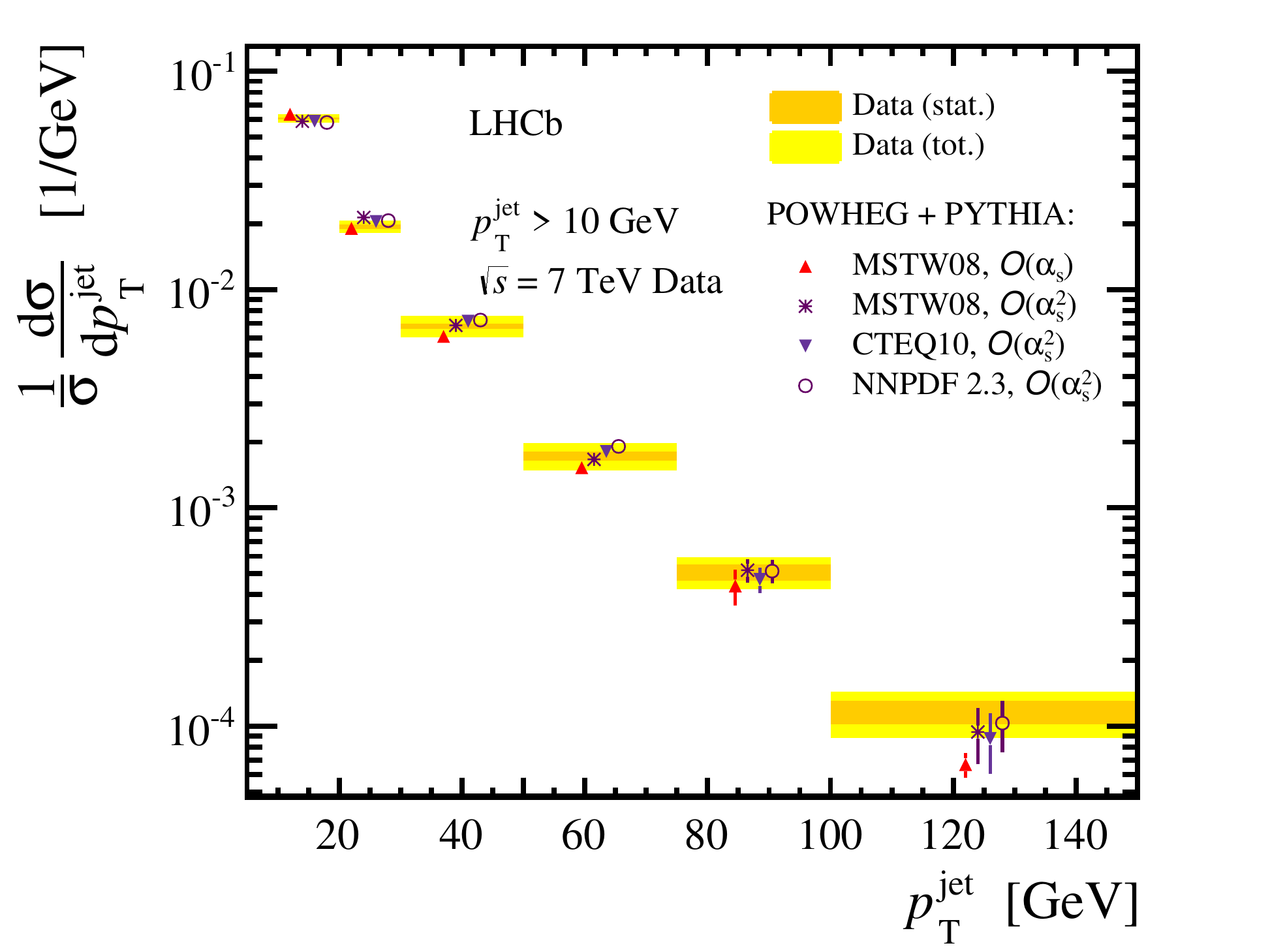}
\hspace{2.cm}
\includegraphics[height=.36\textwidth]{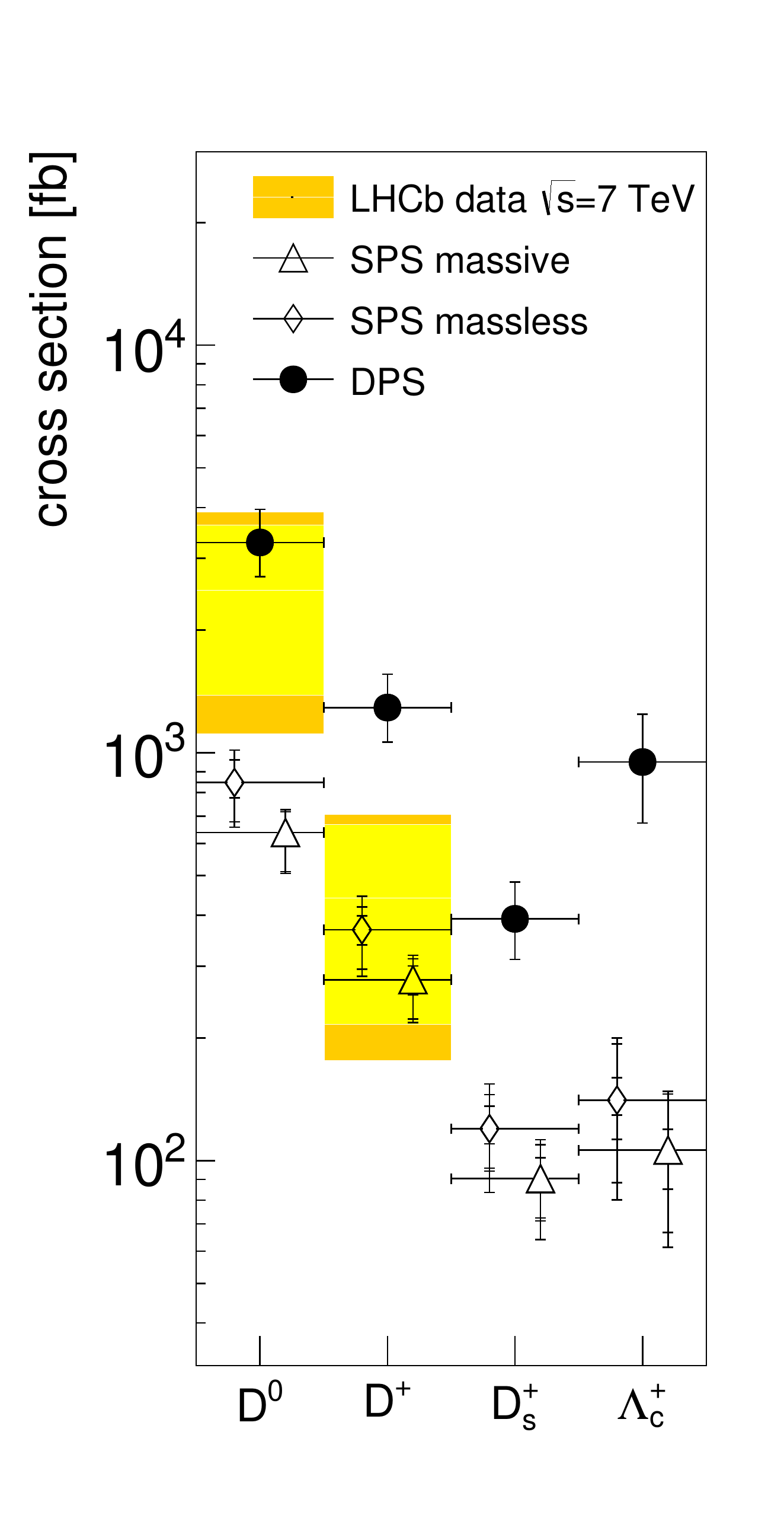}
\hspace{1.cm}
\caption{Associated \Z cross-section measurements with jets~\cite{zmmj} (left) and \D mesons~\cite{zmmd} (right) compared to theoretical predictions.}
\label{fig:CSZa}
\end{center}
\end{figure}

The \W measurements are presented in Figure~\ref{fig:CSW}. All determinations are generally consistent with theoretical predictions calculated at \nnlo in perturbative quantum chromodynamics using the \dynnlo~\cite{dynnlo} generator with six different parameterisations of the proton \pdf. While cross-sections are limited by the knowledge of the integrated luminosity of the data sample, the ratio of the \Wp to \Wm cross-section tests the Standard Model with a precision of about 1.7\%, which is comparable to the uncertainty on the theory predictions.
Figure~\ref{fig:CSZll} (left) shows the inclusive cross-section for \Z production in \pp collisions measured via final states containing charged leptons. Results agree with each other and ratios are found to be consistent with lepton universality. The measurements agree with a \nnlo calculation based on the \fewz~\cite{fewz} generator.
The \Z production cross-section in \pa collisions is measured in both the proton and lead beam directions. Results, which probe the nuclear \pdf in regions where there is no experimental input, agree with predictions and are shown in Figure~\ref{fig:CSZll} (right).

Figure~\ref{fig:CSZa} (left) present the differential cross-section for $\Z+jet$ as a function of the leading jet \pt, not corrected for \fsr of the muons. Reasonable agreement is observed between the Standard Model calculations and the data, where the $\mathcal{O}(\alpha^{2})$ predictions tend to give better agreement with data than the $\mathcal{O}(\alpha)$ calculation.
The cross-section for associated production of a \Z boson with a \D meson is presented in Figure~\ref{fig:CSZa} (right) and compared to theoretical predictions assuming single- or double-parton scattering mechanisms. Results are consistent with expectations for $\Z+\Dz$, while lie below the sum of the two calculations in case of $\Z+\Dp$.

\begin{table}[!t]
\begin{center}
\begin{tabular}{l|lllll}
Process & \multicolumn{5}{c}{Cross-section} \\ \hline
\wpmn & $831$ & $\pm$ $9_{stat}$ & $\pm$ $27_{syst}$ & $\pm$ $29_{lumi}$ & \pb \\
\wmmn & $656$ & $\pm$ $8_{stat}$ & $\pm$ $19_{syst}$ & $\pm$ $23_{lumi}$ & \pb \\ \hline
\zmm & $\phantom{0}76.7$ & $\pm$ $1.7_{stat}$ & $\pm$ $\phantom{0}3.3_{syst}$ & $\pm$ $\phantom{0}2.7_{lumi}$ & \pb \\
\zee & $\phantom{0}76.0$ & $\pm$ $0.8_{stat}$ & $\pm$ $\phantom{0}2.0_{syst}$ & $\pm$ $\phantom{0}2.6_{lumi}$ & \pb \\
\ztt & $\phantom{0}71.4$ & $\pm$ $3.5_{stat}$ & $\pm$ $\phantom{0}2.8_{syst}$ & $\pm$ $\phantom{0}2.5_{lumi}$ & \pb \\ \hline
\zmm in $pA$ & $\phantom{0}13.5$ & $^{+}_{-}$ $^{5.4}_{4.0_{stat}}$ & $\pm$ $\phantom{0}1.2_{syst \oplus lumi}$ & & \nb \\
\zmm in $Ap$ & $\phantom{0}10.7$ & $^{+}_{-}$ $^{8.4}_{5.1_{stat}}$ & $\pm$ $\phantom{0}1.0_{syst \oplus lumi}$ & & \nb \\ \hline
$\zmm+jet_{\pt>10\gevc}$ & $\phantom{0}16.0$ & $\pm$ $0.2_{stat}$ & $\pm$ $\phantom{0}1.2_{syst}$ & $\pm$ $\phantom{0}0.6_{lumi}$ & \pb \\
$\zmm+jet_{\pt>20\gevc}$ & $\phantom{00}6.3$ & $\pm$ $0.1_{stat}$ & $\pm$ $\phantom{0}0.5_{syst}$ & $\pm$ $\phantom{0}0.2_{lumi}$ & \pb \\ \hline
$\zmm+\Dz$ & $\phantom{00}2.50$ & $\pm$ $1.12_{stat}$ & $\pm$ $\phantom{0}0.22_{syst \oplus lumi}$ & & \pb \\
$\zmm+\Dp$ & $\phantom{00}0.44$ & $\pm$ $0.23_{stat}$ & $\pm$ $\phantom{0}0.03_{syst \oplus lumi}$ & & \pb \\
\end{tabular}
\caption{Summary of the cross-section measurements.}
\label{tab:cs}
\end{center}
\end{table}

\begin{thebibliography}{99}

\bibitem{lhcb}
A.~A.~Alves, Jr. {\it et al.} [\lhcb collaboration], JINST {\bf 3}, S08005 (2008).

\bibitem{wz}
R.~Aaij {\it et al.} [\lhcb collaboration], JHEP {\bf 1206}, 058 (2012) [arXiv:1204.1620 [hep-ex]].

\bibitem{zmm}
R.~Aaij {\it et al.} [\lhcb collaboration], LHCb-CONF-2013-007.

\bibitem{zmmpa}
R.~Aaij {\it et al.} [\lhcb collaboration], [arXiv:1406.2885 [hep-ex]].

\bibitem{zee}
R.~Aaij {\it et al.} [\lhcb collaboration], JHEP {\bf 1302}, 106 (2013) [arXiv:1212.4620 [hep-ex]].

\bibitem{ztt}
R.~Aaij {\it et al.} [\lhcb collaboration], JHEP {\bf 1301}, 111 (2013) [arXiv:1210.6289 [hep-ex]].

\bibitem{zmmj}
R.~Aaij {\it et al.} [\lhcb collaboration], JHEP {\bf 1401}, 033 (2014) [arXiv:1310.8197 [hep-ex]].

\bibitem{zmmd}
R.~Aaij {\it et al.} [\lhcb collaboration], JHEP {\bf 1404}, 091 (2014) [arXiv:1401.3245 [hep-ex]].

\bibitem{dynnlo}
S. Catani {\it et al.}, Phys. Rev. Lett. \textbf{98}, 222002 (2007).

\bibitem{fewz}
R.~Gavin {\it et al.}, Comput. Phys. Commun. \textbf{182}, 2388 (2011).

\end{thebibliography}
\end{document}